\documentclass[letterpaper]{JHEP}
\usepackage{epsfig}
\usepackage{latexsym}
\usepackage{amssymb}
\usepackage{amscd}
\usepackage{graphics}
\usepackage{psfig}
\newcommand{\pd}{\partial}

\newcommand{\hp}{{\hat \phi}}
\newcommand{\cF}{{\cal F}}
\newcommand{\cE}{{\cal E}}
\newcommand{\cH}{{\cal H}}

\newcommand{\cO}{{\cal O}}
\newcommand{\hH}{{\hat \cH}}
\newcommand{\sH}{{\ast \cH}}
\newcommand{\hB}{\hat B}

\newcommand{\ag}{g^{{\rm aux}}}
\newcommand{\tr}{{\rho}}

\newcommand{\vx}{{\vec x}}
\newcommand{\vy}{{\vec y}} 
\newcommand{\vq}{{\vec q}}
\newcommand{\vr}{{\vec r}}
\newcommand{\tA}{A^K}

\newcommand{\grad}{\nabla}
\newcommand{\tg}{{\tilde g}}
\newcommand{\cI}{{\cal I}}
\newcommand{\cJ}{{\cal J}}

\newcommand{\qi}{\mathfrak q_1}
\newcommand{\qk}{\mathfrak q_K}
\newcommand{\qv}{\mathfrak q_5}
\newcommand{\q}{\mathfrak q}

\newcommand{\fH}{\mathbb H} 
\newcommand{\fG}{\mathbb G}
 
\newcommand{\fA}{{\mathfrak A}}

\newcommand{\N}{{\bf N}} 

\newcommand{\dl}{{\mathbb D}}
\newcommand{\lo}{{\mathbb L}}
\newcommand{\f}{{\mathfrak f}}
\newcommand{\h}{{\mathfrak h}}
\newcommand{\mmu}{{\mathfrak m}}
\newcommand{\U}{{\cal U}}
\newcommand{\wU}{\widetilde {\U}}

\preprint{SUGP-00/5-1 \\hep-th/0005146}
\keywords{D-branes, Supergravity}
\title{On the Moduli Space of the Localized 1-5 System} 
\author{ Donald Marolf \\ Department of Physics, Syracuse University, \\ 
	Syracuse, NY, USA  \\marolf$@$suhep.phy.yr.edu, \\ }
\author{Sumati Surya\\  TIFR, Homi Bhabha Rd,\\ Mumbai 400 005,
        India. \\ ssurya$@$ tifr.res.in}

\date{June,  2000}
\abstract{We calculate the effective action for small
velocity scattering of localized 1-branes and 
5-branes. Momentum is allowed to flow in the
direction along the 1-branes so that the moduli space has only 1/8 of the
full supersymmetry.  Relative to the more familiar case with the
1-branes delocalized along the 5-branes, this introduces new
moduli associated with the motion of the 1-branes along the 5-branes.  
We consider in detail the moduli space metric for the associated
two body problem.   Even
for motion transverse to the 5-brane, our results differ substantially from
the delocalized case. However, this difference only appears when both the
1-brane charge and the momentum charge are localized.  Despite
the fact that, in a certain sense, 1-branes spontaneously delocalize 
near a 5-brane horizon, the moduli space metric in this limit continues
to differ from the delocalized result.  This fact may be of use in
developing a new description of the associated BPS bound states. 
The new terms depend on the torus size $L$ in such a way that they
give a finite contribution in the $L \rightarrow \infty$ limit.
}

\begin{document}

\section{Introduction}

	 Brane moduli spaces, in particular those of black hole
configurations, have been investigated extensively over the last few years
\cite{fe,shi,khmy,km,mist,gp98,gp99,gp00}.  One motivation is that in
various limits such moduli spaces should be related to the ADS/CFT
conjecture \cite{JM,magoo}.  In particular, recent work
\cite{mist,mpsv00,psv00} has focused on cases connected to the elusive
relation between $1+1$ ADS space and a $0+1$ CFT.

	A closely related motivation of \cite{mist,mpsv00,psv00} is
the attempt to understand the internal states of black holes in terms of the
multi-black hole moduli space.   In contrast to the description 
used in \cite{sv} to account for the entropy of such black holes, this new
description could be valid in a regime of couplings more properly
associated with large classical black holes.  Such a
description could then lead
to better insight into the nature of black hole internal states.

 Since in M-theory the black holes themselves are believed to be marginally
bound states of various types of branes, one might expect the full moduli
space associated with the constituent branes to be relevant to this
problem.  After all, if states in the near horizon limit of the multi-black
hole moduli space are to be interpreted as internal states of a black hole,
then the same interpretation naturally applies to all states in the
near-horizon limit of the moduli space describing the interactions of the
various component branes.  As such, it may be important for this program to
understand any new features inherent in moduli spaces of localized
branes. 
	
	We should stress here that by the term ``localized'' brane we refer
to a p-brane whose {\it classical supergravity} description is in terms of
some p+1 dimensional hypersurface, as opposed to a smeared ``fluid'' of
such objects. It is this set of classical configurations that one would use
to define a moduli space on which one could then consider quantum
mechanical wavefunctions. While such wavefunctions will of course spread
out over the moduli space, this is a different sort of
localization/delocalization than  we will address in this work.

	Now, it is true that when a localized 1-brane approaches a 5-brane
there is a certain `spontaneous delocalization' of the 1-brane charges
\cite{ms98} so that the resulting object resembles the black holes studied
in \cite{mist,mpsv00,psv00}. Thus, one might expect a similar result for
the moduli space dynamics.  However, this is not what we find.  Instead, we
find the moduli space for a localized 1-brane with longitudinal momentum
scattering off such a 5-brane to be substantially different from that of a
delocalized 1-brane in the near-horizon limit.

	With the exception of the test-brane calculations of
\cite{dps,ts97,mal97,chts97}, 1/5-brane moduli space calculations in
supergravity proceed by dimensionally reducing 10 dimensional solutions to
5 or 4 dimensions so that the branes appear as point particles. The
effective action is obtained in the small velocity approximation as shown
by Ferrell and Eardley for Reissner-Nordstrom black holes \cite{fe},
building on previous work \cite{iw,gr}.  The moduli space metric is then
obtained from the kinetic terms, as first demonstrated for BPS monopoles
\cite{manton}.  In several cases, the moduli spaces have been related to
the target space of 1-dimensional supersymmetric sigma models and a
connection made between the number of supersymmetries of the effective
theory and the complex structure on the moduli space
\cite{gps97,gp98,gp00}.

   	In order to get an effective theory of point particles, the brane
configuration in 10-dimensions must have appropriate isometries along all
possible brane directions. If however, one begins with both 5-branes and
localized 1-branes, the 1-branes break translational symmetry along the
5-branes and dimensional reduction in these directions is not possible.
The effective theory is necessarily one of extended objects.  Localized
brane configurations have received considerable interest in recent years,
but a study of the scattering of such objects has not yet been carried out.
It is the purpose of this work to extend the calculations of
\cite{fe,km,mist,gp99} to a particular localized system.

	   In this paper, we calculate the moduli space metric for the
system of \cite{ms98} containing Neveu-Schwarz 5-branes and localized
fundamental strings (F1-branes) in the near horizon limit.  By S-duality,
the moduli space of the localized D1/D5-brane system with momentum will be
identical.  The solution has a 5-brane wrapped on a $T^5$ and a separated
1-brane wrapped on one of the $T^5$ cycles.  Thus, unlike the delocalized
case, there are four extra moduli in the problem labeling the location of
the string in the remaining $T^4$ directions. This means that there is a
single spatial isometry along the cycle on which the string is wrapped and
along which a dimensional reduction can be performed. We also include a
third charge corresponding to momentum directed along the string. We then
calculate the effective action in the low velocity limit following Ferrell
and Eardley \cite{fe}.

The procedure for computing the moduli space involves first replacing the
branes with a smooth `dust' source and then taking the distributional limit
where the dust describes a set of branes.  We derive the effective action
for smooth dust sources in section 2.  One of the nice features of this
approach is that it is insensitive to the details of the 1-brane
singularity.  Any distribution of 1-brane charge localized in a region much
smaller than the length scales $L$ and $r_5$ associated with the size of
the 4-torus and the 5-brane charge, respectively, will produce much the
same results.  For large $L$, $r_5$ the curvatures and dilaton remain small
at the 1-brane source, and ten-dimensional supergravity is an adequate
description of the system.

The action derived in section 2 can be readily generalized to include many
independent dust distributions. In section 3, we consider a special two
body case describing a single localized stack of 1-branes and a single
stack of 5-branes carrying delocalized 1-brane and momentum charges.  We
obtain an explicit expression for the moduli space metric in the limit in
which the localized branes approach the 5-brane horizon.  As with the
string metric for a single 5-brane, the metric has a warped product
structure, with the transverse radial directions warping the internal $T^4$
directions.  As mentioned above, the metric in the transverse directions
differs from that of the delocalized case.  In particular, the transverse
metric for relative motion in isotropic coordinates is no longer simply a
conformal factor times the standard Euclidean metric.  In addition to the
terms familiar from the delocalized case, we identify a new one which
depends on the ratio $r_5/L$.  This term is sufficiently large for large
$L$ to make a finite contribution in the limit $L \rightarrow \infty.$ We
close with some discussion in section \ref{disc}.

\section{The Effective Action} 

In the string frame, the ten dimensional action of type IIB supergravity
contains the terms
\begin{equation} 
\label{IIBaction}
S_{10}={1 \over 16 \pi G_{10}} \int d^{10}x  \sqrt{-G}e^{-\hp}[ ^GR+ (\pd
\hp)^2  -{1\over 12} {{\hat \cH}}^2], 
\end{equation} 
where $G_{MN}$ is the 10 dimensional string metric, $\hp$ is the dilaton
and $\hH$ is the 3-form field associated with an antisymmetric
Neveu-Schwarz 2-form field, $\hB$.  The symbol ${}^GR$ refers to
the Ricci scalar of $G_{MN}$ as opposed to that of other metrics that
will appear later.  A stationary point of (\ref{IIBaction})
represents a solution of type IIB supergravity with all fermions and
Ramond-Ramond fields set to zero.  The hats on fields serve to simplify the
notation later in the paper, after we dimensionally reduce the solution
along the single translational symmetry.  The overall normalization of the
action will not be needed for our purposes.

	We are interested in the case of a separated F1-NS5 brane solution,
where the one brane is localized in the transverse 5-brane directions.
Such a solution was found in \cite{ms98}, and belongs to the class of
chiral null models of \cite{ht}. Five of the spatial directions are
compactified on a $T^5$ on which the 5-brane is wrapped. The 1-brane is
wrapped along a single cycle of the $T^5$. We employ the coordinates
$(t,z,x^i,y^a)$, where $z$ is the direction along which the 1-brane is
wrapped, $x^i$ are the 4 spatial directions transverse to the 5-brane, and
$y^a$ are the remaining 4 directions transverse to the 1-brane along the
5-brane.  For simplicity we take the $T^5$ to be an orthogonal torus with
$z,y^a$ labeling the orthogonal directions and with the corresponding
cycles having
length $L_z,L$.  In these coordinates the non-vanishing components of the solution
are,
\begin{eqnarray}
\label{cnm.eq} 
G_{MN}dX^MdX^N &=& H_1^{-1}du( -dv + Kdu)+H_5 dx^i dx^i + dy^ady^a
\nonumber \\
e^{-\hp} &=& {H_1\over H_5} \nonumber \\ 
\hH_{ijk} &=& -\epsilon_{ijkl}\partial_l H_5 \qquad \hB_{uv}=G_{uv} 
\end{eqnarray} 
where, $H_5(x^i), H_1(x^i,y^a),K(x^i,y^a)$ are functions associated
respectively with
the 5-brane charge, the 1-brane charge, and momentum in the z-direction.
{}From \cite{ht}, it follows that they satisfy the coupled
equations:   
\begin{eqnarray} 
\label{H1.eq}
\pd_i^2 H_5 &= & -c_5\tr_5 \nonumber \\   
\pd_i^2 H_1 +H_5\pd_a^2H_1&=& - c_1\tr_1 \nonumber  \\
\pd_i^2 K +H_5\pd_a^2K&= & -  c_K\tr_K,  
\end{eqnarray} 
where $\tr_5,\tr_1,\tr_K$ are the brane charge densities for the
5-brane charge, 1-brane charge and momentum respectively.  The
solution for a 1-brane separated in the transverse
directions from a 5-brane was found in \cite{ms98} (see
Eq. (\ref{solution.eq})).  Being a chiral null solution, it preserves $1/8$
of the supersymmetries, or 4 supercharges. Unlike the solutions considered
earlier \cite{km,mist,gp99,gp00}, neither $H_1$ nor $K$ are harmonic
functions, which is the crucial point of departure for this analysis. 

	Our choice of convention will be to take $x^i$ along the
$(1,2,3,4)$ directions and the $y^a$ along the $(5,6,7,8)$ directions. In
what follows, we use the collective spatial label $\alpha=(i,a)$.

	The isometry along the $z$ direction makes it possible to 
dimensionally reduce this solution to 8+1  spacetime
dimensions. Proceeding as in \cite{mahsch}, we find the 
8+1-d  non-vanishing fields $g_{\mu\nu}, A^K_\mu, A^1_\mu, \phi, \psi, 
{\cal H}_{\mu_1 \mu_2 \mu_3}$,
\begin{eqnarray}
\label{9dsol.eq}
ds^2_{8+1} &=& g_{\mu\nu} dx^\mu dx^\nu \nonumber \\	     
&= & - (H_1H_K)^{-1} dt^2 + H_5 dx^2 + dy^2 \nonumber \\	     
\tA_0 &=& G_{0 z}={H_K-1 \over H_K} \nonumber \\
A^1_0 &= & \hB_{0z}={1\over H_1} \nonumber \\  
\phi &= &\hp-{1\over 2}\ln G_{zz}=\ln {H_5\over \sqrt{H_1H_k}} \nonumber \\ 
\psi &=& \ln G_{zz}= \ln {H_K\over H_1} \nonumber \\
\cH_{ijk}&=&\hH_{ijk}= -\epsilon_{ijkl}\partial_l H_5, 
\end{eqnarray}  
where $H_K=1+K$, $dx^2 = \sum_{i=1}^4 dx^idx^i$, and $dy^2 = \sum_{a=1}^4
dy^a dy^a$.  The notation reflects the fact that
the potential $A^1_\mu$ couples to the 1-brane charge while 
$A^K_\mu$ couples to momentum.

As we can see from (\ref{9dsol.eq}), the 5-brane couples magnetically
to the field strength ${\cal H}$.  However, in order to explicitly
couple the potentials to sources, it is 
useful to work in a formalism where the charges all couple
electrically to the gauge fields.  Now that we have reduced
the system to 8+1 dimensions, the 3-form field strength produced by the
5-brane does not couple directly to any other charges.  Thus, we
are free to consider a dual 6-form
field strength and the associated potential.  One can check that, in order
for the dual 6-form field strength to be an exact form, one must take the 
dual using the
auxiliary metric  $\ag_{\mu\nu}=
(H_1H_K)^{1\over 3} H_5^{-{2\over 3}} g_{\mu\nu}$,
\begin{equation} 
\sH_{\mu_1 \mu_2\mu_3\mu_4\mu_5 \mu_6}={1\over \sqrt{-\ag}}\quad \epsilon_{\mu_1 \mu_2 \mu_3\mu_4\mu_5
\mu_6}^{({\rm aux})\quad \qquad\nu_1\nu_2\nu_3} \cH_{\nu_1\nu_2\nu_3}.
\end{equation} 
For the solution (\ref{9dsol.eq}), the associated potential $A^5_{\mu_1 \mu_2
\mu_3 \mu_4 \mu_5}$ for $\sH$ (i.e., satisfying $dA^5 = \sH$) 
then has a single nonzero component, 
\begin{equation} 
A^5_{05678}=H_5^{-1}.
\end{equation} 
Finally, since our 5-brane will always remain parallel to the 5,6,7,8 directions
(and to the 9 direction, which is hidden in the 8+1 formalism), it is
convenient to introduce the notation
$A^5_\mu = A^5_{\mu5678}$, so that $A^5$ can be described as a 
vector potential in parallel with $A^1$ and $A^K$. Note, however, that
the 5,6,7,8 components of $A_\mu$ will always vanish.  For
the static solution we have      
\begin{equation} 
A^5_{0}=H_5^{-1}.
\end{equation}

The dimensionally reduced 8+1 dimensional action is, 
\begin{eqnarray} 
\label{888}
S &= &{1 \over 16 \pi G_9} \int d^{9} x {\sqrt {-g}}e^{-\phi}[ R+\pd^\mu \phi \pd_\mu \phi
- {1\over 4} \pd^\mu \psi \pd_\mu \psi  -{1 \over
4}e^{-\psi}\cF^{\mu\nu}\cF_{\mu\nu} \nonumber \\ 
&& -{1 \over 4}e^{\psi}\cE^{\mu\nu}\cE_{\mu\nu}  -{1\over 2 \times
6!}e^{2\phi} \sH^2] + {1\over 4}\int \sH \wedge \cF \wedge A^K +S_{{\rm
source}}, 
\end{eqnarray} 
where $\cF,\cE$ are the field strengths associated with the fields, $A^1_\mu$
and $A^K_\mu$ respectively and $G_9 = G_{10}/L_z$. 
The source terms are $S_{source}= S_{matter}
+S_{current}$.  The first of these contains the kinetic terms for the branes:
\begin{eqnarray} 
\label{matter}
S_{matter} &= & -{1 \over 16 \pi G_9} \int dt d^4xd^4y \{-c_1 \tr_1{\sqrt{{-d\tau_1^2\over dt^2}}}
e^{{\psi \over 2}}  -c_K \tr_K{\sqrt{{-d\tau_K^2\over dt^2}}}
e^{-{\psi \over 2}} \nonumber \\ && 
-c_5 \tr_5{\sqrt{{-d\tau_5^2\over dt^2}}} e^{-\phi}], 
\end{eqnarray}
where the 5-brane kinetic term takes a form like that of a point particle
due to our condition that the 5-brane remain parallel to the $y^a$
directions. Here $\tau_1,\tau_K, \tau_5$ denote the proper time measured
along the various branes. 
The current term is
\begin{eqnarray}
S_{current}&= & -\int dt d^4xd^4y [c_1\tr_1 A^1_\mu v_1^\mu+ c_K\tr_K A^K_\mu
v_K^\mu + c_5\tr_5 A^5_\mu v_5^\mu ], 
\end{eqnarray} 
where we have taken the matter to be pressureless dust as in \cite{fe},
with $v_1^\mu$, $v_K^\mu$, and $v_5^\mu$ the velocities of the 1-brane, 
momentum, and 5-brane charge distributions (`dust')
respectively.   We take these velocities to be functions of $t$ only.
Since the $y^a$ (i.e., 5,6,7,8) components of $A^5_\mu$ always vanish, 
the corresponding components of $v_5^\mu$ are irrelevant.  This is consistent
with the fact that only the velocity of the 5-brane in the directions
transverse to its world-volume are well-defined.  We will continue
to represent this velocity as a 9-vector, following our notation for
$v_1^\mu$ and   
$v_K^\mu$, but with the understanding that we set the $y^a$ components
of $v^\mu_5$ to zero.

In order to regularize the solution, we take $\tr_I$ (for $I =1,K,5$)
to be a smooth function.  
In the case of $\tr_5$, the density
will be translationally invariant in the torus directions ($y^a$).  
The limit of localized brane sources leads to the known static
solutions. We follow the approach of \cite{fe} in first deriving the
effective action for smooth sources in the slow motion approximation and
then taking the limit of localized brane sources. 
	
	These matter sources can be justified either by arguing (as in
\cite{fe}) that any smooth source should be able to approximate a black
hole, or by noting that $S_{matter}+ S_{current}$ follows from the relevant
parts of the Dirac-Born-Infeld and Wess-Zumino terms in the brane effective
actions (see, e.g. \cite{pol}). Viewed in this second way, it is a part of our
ansatz that the internal gauge fields are set to zero. 

Due to the BPS nature of the branes, the solution (\ref{cnm.eq}) is static.
The moduli for this system are just the brane's spatial locations.
To calculate the metric on this moduli space  we consider the small
velocity approximation of \cite{fe,km} in which the forces between the
branes remain small. The motion of the branes in this approximation is
along geodesics on this moduli space so that its  metric can
simply be read off from the effective action.
  
As in \cite{fe}, the time reversal symmetry\footnote{The action (\ref{888})
with the Chern-Simons term does not have time reversal symmetry, but the
original action (\ref{IIBaction}) does have this symmetry. Thus, the
equations of motion for the physical fields must be time reversal
invariant. } can be used to argue that to first order in velocities the
perturbations take the form:
\begin{eqnarray} 
g_{\mu\nu}dX^\mu dX^\nu &= & - H_1^{-1}H_K^{-1}dt^2 +\delta_{ij}H_5^{-1} [dx^i + N^idt][dx^j
+N^jdt] \nonumber \\ && +\delta_{ab}[dy^a +N^adt][dy^b +N^bdt] \nonumber \\
\delta A^1_\alpha& = & A^1_\alpha \nonumber \\
\delta \tA_\alpha& = & \tA_\alpha \nonumber \\
\delta A^5_{\alpha }& = & A^5_\alpha, \nonumber \\ 
\end{eqnarray} 
where $\alpha$ runs over the spatial directions. Note that $H_1,H_K,H_5$
are now time dependent since the sources in (\ref{H1.eq}) are time
dependent. The perturbation in the metric appears as a non-vanishing shift
$N_\alpha$.

The next step is to compute the $O(v^2)$ effective action.
As in \cite{fe}, one can show that only those $O(v^2)$ terms which
follow from the above $O(v)$ expansion of the fields will 
in fact contribute to the equations of motion.  
Other $O(v^2)$ terms in the action do not contribute 
due to the fact that we are near a stationary point of the full action.
For this reason, we include below only terms that arise from first order
variations in the fields.

Note that
an 8+1 split of the spacetime into time and space is inherent in 
the slow motion approximation.  As a result, the fields below will
be written with the index $\alpha$ that
runs only over spatial directions.  It is convenient at this point to
make a change of conformal frame and to introduce a rescaled metric
\begin{equation}
\label{resmetric}
d\tilde s_8^2 = \tilde g_{\alpha \beta} dx^\alpha dx^\beta = H_5^{-1}
g_{\alpha \beta} dx^\alpha dx^\beta = dx^2 + H_5^{-1} dy^2. 
\end{equation}
The spatial indices $\alpha, \beta$ will be raised and lowered with
the rescaled metric (\ref{resmetric}).

In introducing this convention, it is important to point out that
indices will arise in only two ways.  One class of indices come
from differential forms such as $A^1_\alpha$, and
$\partial_\alpha H_1$.  In these cases, a covariant placement of the indices
is natural and the objects with lower indices $\alpha, \beta$, etc.
are simply the pull-back of the spacetime objects (with lower indices $\mu,
\nu$, etc.) to the spatial slice.  All other indices appear on the
velocities $v_I^\mu$.  For such objects, a contravariant placement of
the indices is natural and $v_I^\alpha$ represents simply the restriction
to the set of spatial components.  In contrast, when applying our 8+1
decomposition in the conformal frame (\ref{resmetric}) to an expression
involving $A^\mu$ or $v_\mu$, one must think carefully about the factors
of $H_5$.  Despite this initial complication, the rescaled metric
(\ref{resmetric}) simplifies the results sufficiently as to make its
introduction worthwhile.

A long calculation leads to the $O(v^2) $ effective action,
\begin{eqnarray} 
\label{long}
S &= & {1 \over 16 \pi G_5 L^4} \int dt d^4xd^4y \Biggl[ 
 - \sum_{I\in \{1,K,5\}} \tr_I- H_K{\dot H_1}{\dot H_5}-H_1{\dot H_K}{\dot
H_5}-H_5{\dot H_1}{\dot H_K}   \nonumber \\ 
&-& {1 \over  H_5\mu_\N}{dP^\N \cdot dP^\N\over2} 
+ \sum_{I\in \{1,K,5\}} \Biggl( 
{c_I\tr_I \mu_I H_5\over 2} v_I^\alpha v_{I\alpha} \nonumber \\ 
&-&  {1\over { 2 H_5 \mu_I}} \left({H_I dP^I \cdot 
dP^I\over2}  - dP^I \cdot dP^\N \right)  
+  P^I_{\alpha}\left[ -\pd_t \left( \tilde g^{\alpha \beta} \pd_\beta H_I 
\right) +c_I\tr_I v_I^{\alpha} \right]  \Biggr) 
 \nonumber \\ &+& \sum_{I \in \{1,K,5\}}{1\over 4 \mu_I H_5} \Biggl(\sum_{J
 \in \{1,K,5\};  J\neq I} H_J{\epsilon_{\alpha \beta \gamma \delta }(dP^I_{\alpha \beta}
dP^J_{\gamma \delta} )\over 2} - 
 \epsilon_{\alpha \beta \gamma \delta} (dP^N_{\alpha \beta} dP^I_{\gamma \delta}) 
\Biggr)\nonumber \\ &+& {1\over \mu_N
H_5}{\epsilon_{\alpha \beta \gamma \delta}(dP^N_{\alpha \beta} dP^N_{\gamma \delta}
)\over 4}  \Biggr],  
\nonumber \\
 \end{eqnarray} 
where we have defined
\begin{eqnarray}
\mu_I  &=& {{H_1H_K} \over {H_I}}, \ \ \ {\rm for} \ I \in \{1,K,5\}
\nonumber \\
\mu_\N &=& H_1 H_K, \nonumber \\ 
\epsilon^{\alpha\beta\gamma\delta}&=&\epsilon^{0 \alpha
\beta\gamma\delta 5678}, 
\end{eqnarray}
and we have introduced the one-form fields
\begin{eqnarray} 
P^I_{\alpha} &= &  A^I_\alpha + \mu_I N_\alpha, \ \ \ {\rm for} \ 
I \in \{1,K,5\}, \nonumber \\
P^\N_{\alpha}&=& \mu_\N N_\alpha,   
\end{eqnarray} 
with $dP^I_{\alpha\beta}=\pd_\alpha P^{I}_\beta -\pd_\beta
P^{I}_\alpha$, and $dP^{I}\cdot dP^{I} = \tilde g^{\alpha \gamma}
\tilde g^{\beta \delta}
dP^{I}_{\alpha \beta} dP^I_{\gamma \delta}$, etc.

In order to obtain the moduli space metric from this reduced action we need
to be able to write down the velocity dependence explicitly. As described
in \cite{fe}, the values of $P^I,P^\N$ are to be determined
from the constraints, which are in fact given by varying (\ref{long})
with respect to $P^I,P^\N$. The
existence in the delocalized case of a simple solution to these equations
in terms of $\tr_I$ and $v_I^\alpha$ \cite{gp99} is rather
special.

It is not at all obvious that the same should be true for the localized
case.  From (\ref{long}), we see that  the equations of motion for the
localized case are  
\begin{eqnarray}
\label{Ieqn}
&& \partial_\alpha \left( {1 \over {H_5 \mu_I}} [ H_I dP^{I \alpha \beta}
- dP^{\N \alpha \beta}]  -{\epsilon^{\alpha\beta \gamma\delta}\over
2H_5\mu_I} [\sum_{J \in \{1,K,5\}; J\neq I} H_JdP^J_{\gamma\delta}
-dP^N_{\gamma\delta}]\right)  \nonumber \\  &&= 
\pd_t \left( \tilde g^{\alpha \beta} \pd_\alpha H_I 
\right) -c_I\tr_I v_I^{\beta}, \nonumber  \\
&& \\
&&\partial_\alpha \left( {1 \over {H_5}} [ {{2 dP^{\N \alpha \beta}} \over {
\mu_\N}}
- \sum_{I \in \{1,K,5\}} {{dP^{I \alpha \beta}} \over {\mu_I}}
] -{\epsilon^{\alpha\beta \gamma\delta}\over
2H_5} [{2 dP^N_{\gamma \delta} \over \mu_\N} -\sum_{I \in \{1,K,5\}}
{dP^I_{\gamma\delta} \over \mu_I}]\right)  \nonumber \\ 
&&=0.
\end{eqnarray}

The trick to solving such equations is of course to first write the
right hand side as the divergence of some antisymmetric tensor.
For (\ref{Ieqn}) for the case $I=5$ this is straightforward and proceeds
along the lines of \cite{fe}.
One first notes that the right hand side is nonzero only when the index
$\beta$ takes values in the transverse directions ($\beta = i$).
One then uses the constraint equation (\ref{H1.eq}) for $H_5$.
By combining this constraint with current conservation, one arrives 
at a conservation equation for $H_5$ itself:
\begin{equation}
\label{H5cons}
\dot{H}_5 + v_5^i \partial_i H_5 =0.
\end{equation}
By using this result, and also using the constraint for $H_5$ to
express $\tr_5$ in terms of $H_5$, one can express the
right hand side of (\ref{Ieqn}) for $I=5$ as 
$ \partial_j L^{5ji}$ where
\begin{equation}
\label{5pot}
L^{5ji} =  \tg^{jk} \partial_k  H_5 v_5^i  - \tg^{ik} \partial_k
H_5 v_5^j.  
\end{equation}

For $I=1,K$ the equations are more complicated.  If one tries to again follow
\cite{fe}, the root of the problem is that $H_1$ and $H_K$ are coupled
to $H_5$ through the constraints (\ref{H1.eq}).  Thus, even if 
the 1-brane isn't moving,   the
field $H_1$ at a given point will change if we move a 5-brane.
The result is that $H_1$ and $H_K$ do not satisfy simple conservation
equations of the form (\ref{H5cons}).

Nonetheless, one can make progress by introducing a few more
potentials.  Let us first generalize (\ref{5pot}) to $I=1,K$ and to
an antisymmetric tensor on the full space (thus defining the 
$y^a$ components)   
through:
\begin{equation}
\label{kij.eq}
L^{I\alpha \beta} = \tg^{\alpha \gamma} \partial_\gamma H_I v_I^\beta -
\tg^{\beta \gamma} \partial_\gamma H_I v_I^\alpha 
\end{equation}
Note that for $I=5$, equation
(\ref{5pot}) is reproduced  with $L^{5ai}=L^{5ab}=0$.
In addition to (\ref{kij.eq}), we also need the extra potentials
\begin{equation} 
\cJ^I  \equiv \grad^{-2}(\pd_t +v_I^\alpha\pd_\alpha)(H_IH_5), 
\end{equation} 
where $\nabla^2 = \sum_i \partial_i \partial_i.$
Note that $\cJ^5=0$ due to the conservation law (\ref{H5cons}).
It is useful to associate with $\cJ^I$ a set of antisymmetric tensor fields 
$\cJ^I_{\alpha \beta}$ defined by:
\begin{eqnarray}
\cJ^I_{ia} \equiv & \partial_i \partial_a \cJ & \equiv 
- \cJ^I_{ai} \nonumber \\   
\cJ^I_{ab} \equiv  & 0 & \equiv \cJ^I_{ij}.
\end{eqnarray}
A bit of calculation then shows that 
the right hand side  of (\ref{Ieqn}) may be cast
in the form,  
\begin{equation} 
\pd_t \left( \tg^{\alpha \beta} \pd_\alpha H_I 
\right) -c_I\tr_I v_I^{\beta} = \partial_\alpha 
(L^{I\alpha \beta } + {1\over H_5}\cJ^{I\alpha \beta }).
\end{equation} 
This allows us to obtain the following solutions: 
\begin{eqnarray}
\label{pqsols.eq}
2(1-{\epsilon \over 2})\cdot {dP^N \over \mu_\N} &=& (1-{\epsilon \over
2})\cdot \sum_{I
\in{1,K,5}} {dP^I \over \mu_I} \\ 
(1+{\epsilon \over 2})\cdot dP^I &=& (1+ {\epsilon \over 2})\cdot [{\mu_I
H_5 \over H_I}L^I -{1\over H_I} \sum_{J\in {1,K,5}}{H_5 \mu_\N \over
H_J}L^J],   
\end{eqnarray} 
where we have used the notation
\begin{equation}
(1 \pm {\epsilon \over 2})^{\alpha \beta \gamma \delta} = 
(\tilde g^{\gamma [\alpha} \tilde g^{\beta]\delta} \pm {1\over
2}\epsilon^{\alpha \beta \gamma \delta} ),
\end{equation} 
and
\begin{equation}
\left[\left( 1 + \frac{\epsilon}{2}\right) \cdot A \right] =
\left( 1 + \frac{\epsilon}{2}\right)_{\alpha \beta}^{\gamma \delta}  A_{\gamma \delta}
\end{equation}
for an antisymmetric tensor $A$.
Using the fact that $\cJ_{ij}=0=\cJ_{ab}$, one can check that when the
5-brane charge vanishes, the $(ij)$ and $(ab)$ components reduce to the
results of \cite{khmy}.  The appearance of $\cJ_{ai}$ in the $(ia)$
components is a novel feature of our calculation.

Inserting the above results into the effective action  yields:
\begin{eqnarray} 
\label{EA}
S &= & {1 \over 16\pi G_5 L^4}\int dt d^4xd^4y \Biggl[-  \sum_{I\in
\{1,K,5\}} \tr_I - H_K{\dot H_1}{\dot H_5}-H_1{\dot H_K}{\dot H_5}-H_5{\dot
H_1}{\dot H_K}   \nonumber \\  
&+& 
\sum_{I\in \{1,K,5\}} \left( 
{c_I\tr_I \mu_I H_5\over 2} v_I^\alpha v_{I\alpha}  \right) 
- \frac{H_1}{2}(1+{\epsilon \over 2})^{\alpha \beta \gamma \delta}  
(L^K + {1 \over H_5}\cJ^K )_{\alpha \beta} (L^5 + {1 \over
H_5}\cJ^5)_{\gamma \delta} \nonumber \\ &-&   
 \frac{H_K}{2} (1+{\epsilon \over 2})^{\alpha \beta \gamma \delta}
(L^5 + {1 \over H_5}\cJ^5)_{\alpha 
\beta} (L^1 +{1 \over H_5} \cJ^1)_{\gamma \delta} \nonumber \\ &-&
\frac{H_5}{2}(1+{\epsilon \over 2})^{\alpha \beta \gamma \delta}
(L^1 + {1 \over H_5}\cJ^1)_{\alpha
\beta} (L^K + {1 \over H_5}\cJ^K)_{\gamma \delta}  \Biggr], 
\end{eqnarray}
where $G_5 = G_9/L^4.$

Up until this point, for each type of charge $I=1,K,5,$
we have allowed for only one dust distribution $\tr_I$ with a single
constant velocity $v_I^\mu$.  However, the form (\ref{EA}) provides
a ready generalization to the case of many independent
dust distributions $\tr^A_I$ (representing a different stack of branes
for each value of $A$) with independent velocities $v_I^{A\mu}$ for $I=1,K$.
Note that we may write $H_I = 1 + \sum_A \tilde H_I^A$ where
$\tilde H_I^A$ satisfies equation (\ref{H1.eq})  for the source
$\tr^A_I$ and vanishes at infinity.
The linearity of the constraint equations and of the equations of motion
for $dP^I,dP^\N$ then imply that the effective action for the
multi-brane case is again of the form (\ref{EA}) with a separate
kinetic term (${c_I\tr_I^A \mu_I H_5\over 2} v_I^{A\alpha} v^A_{I\alpha}
$) included for each brane and with $L^I_{\alpha \beta}$ and
$\cJ^I$ given by: 
\begin{eqnarray}
\label{multi1}
L^{I\alpha \beta} = \sum_A \left(
\tg^{\alpha \gamma} \partial_\gamma \tilde H_I^A
v^{A\beta} -
\tg^{\beta \gamma} \partial_\gamma \tilde H_I^A v^{A\alpha}
\right) \nonumber \\
\cJ^I  \equiv \sum_A \grad^{-2}(\pd_t +v_I^{Ai}\pd_i +v_I^{Ai} \pd_a)(\tilde
H_I^A H_5).
\end{eqnarray}   

The structure here is similar to that of the delocalized case, with the
main new feature being the terms of the form $\cJ_{\alpha \beta}$.

\section{The Two-Body Problem} 

Our task now is to take a limit in which the smooth dust sources become
distributions representing some set of localized branes and to then
evaluate the effective action (\ref{EA}).  The result should yield an
action quadratic in velocities associated with geodesic motion through some
moduli space. 

As one might expect, the fully general case for localized branes
is quite complicated.  We therefore pick out a special two-body case
for detailed analysis.  Two-body problems are particularly simple due
to the symmetry about the axis connecting the two bodies.  This symmetry
causes many terms to vanish, and the resulting effective action takes a
tractable form.  In particular, no term involving $\epsilon^{\alpha \beta \gamma \delta}$
in (\ref{EA}) will contribute in this case.  To see this, note that since
${\cal J}^I_{ij}=0$ we have $\epsilon \cdot {\cal J}^I = 0$.  As a result, 
(\ref{EA}) shows that $\epsilon$ always appears in the combination $\epsilon_{\alpha \beta
\gamma \delta} v^\gamma u^\delta$ where $v^\gamma$, $u^\delta$ are the velocities of the
two objects.  By Galilean invariance, it is sufficient to note that such terms
vanish in the center of mass frame where $v^\gamma$ and $u^\delta$ are proportional.

\subsection{The setting}

Our original goal was to study the scattering of localized 1-branes and
5-branes.  As noted above, an object cannot simultaneously carry localized
1-brane charge and 5-brane charge \cite{ms98}.  For this reason, we take
one of our two objects
to be a stack of localized 1-branes carrying some longitudinal momentum, with
$v^\alpha$ denoting the velocity of this object.  We take the other object
to be a stack of 5-branes, which is also allowed to carry (delocalized)
1-brane and momentum (K) charges.  We denote the velocity of this object by
$u^\alpha$.  Note that the velocity components $u^a$ of such an object
around the torus are ill-defined.  It is consistent to set them to zero,
and we do so for convenience.  We refer to the two objects as the
localized object ($\lo$) and the delocalized object ($\dl$), where as usual
`delocalized' means delocalized along the torus directions.  

It is useful to decompose the various $H_I$ into parts corresponding to
the two objects:
\begin{equation} 
H_I= 1+ H_I^\dl +H_I^\lo,  
\end{equation} 
where the $H_I^{\lo,\dl}$ have corresponding sources $\tr^\dl,\tr^\lo$,
and vanish  at infinity.  Notice that $H_5^\lo =0$.  
Since the delocalized part is translationally invariant in the $y^a$
directions, it satisfies the constraint
\begin{equation} 
c_I\tr_I^\dl =  - \nabla^2 H_I^\dl,
\end{equation} 
which is independent of $H_5$. It therefore obeys the the conservation law
(\ref{H5cons}),
\begin{equation}
\label{Hdelcons}
\dot{H}_I^\dl + u^i \partial_i H_I^{\dl} =0,
\end{equation}
and so does not contribute to the potentials $\cJ^I$.
 
It then follows from (\ref{multi1}) that we have the relations
\begin{eqnarray}
\label{multi}
L^{I  \alpha \beta} &= & \tg^{\alpha \gamma} \partial_\gamma 
H_I^\dl u^{\beta} - \tg^{\beta \gamma} \partial_\gamma H_I^\dl
u^{\alpha} \nonumber \\ 
&& + \tg^{\alpha \gamma} \partial_\gamma 
H_I^\lo v^{\beta} - \tg^{\beta \gamma} \partial_\gamma H_I^\lo
v^{\alpha} \nonumber \\ 
\cJ^I  &=& \grad^{-2}(\pd_t +v^{\alpha}\pd_\alpha)( H_I^\lo H_5). 
\end{eqnarray}

After performing several integrations by parts we find the effective
action to be
\begin{eqnarray} 
\label{sr.eq}
S_{eff} &= &{1 \over 16\pi G_5 L^4} \int dt d^4x d^4y \Biggr[  -  \sum_{I\in \{1,K,5\}} \tr_I + {1\over2}[v^iv^i +v^av^a] \left( 
c_1\tr_1^\lo
+c_K\tr_K^\lo \right) \nonumber \\ 
&+& {1 \over 2} u^iu^i\left(c_5\tr_5 + c_1\tr_1^\dl +c_K\tr_K^\dl \right)
+ (u^i-v^i)(u^i-v^i) {c_5\tr_5 \over 2}(H_1^\lo + H_K^\lo +H_1^\lo
H_K^\lo) \nonumber \\ &+& {1\over 2} \left[ (u^i - v^i)(u^i - v^i)H_5 + v^av^a \right ] \left(
c_1\tr_1^\lo H_K^\dl  + c_K\tr_K^\lo H_1^\dl \right) \nonumber \\
&+&  
\frac{1}{2}[(\partial_t + v^i \partial_i)
H_5][(\partial_t + v^i \partial_i)(H_1^\lo H_K^\lo)]
\Biggl].  
\end{eqnarray} 
In the expression above, a sum over $i,j,a$ is implicit.

As will become evident in what follows, the key feature of this action is
the last term.  This term turns out to be rather subtle.  Note, however,
that it would vanish if $H_1^\lo$ and $H_K^\lo$ were homogeneous on the
torus, as in that case $H_1^\lo$ and $H_K^\lo$ each satisfy $(\partial_t +
v^i\partial_i) H^\lo_{1,K}=0.$ The integral over the torus in fact allows
us to replace both $H_1^\lo$ and $H_K^\lo$ in this term by $\hat{H_1^\lo}$ and
$\hat{H_K^\lo}$, where the hat indicates that we have removed the
homogeneous mode from the Fourier expansion of each function on the torus.
It will turn out to be important to note this explicitly.  The reason is
that, in order to evaluate this final piece in terms of the sources, we
will need to write it without explicit time derivatives.  In fact, the
constraints and the `conservation' of $H_5$ can be used to write this last
term in the form:
\begin{equation}
\frac{1}{2}[(\partial_t + v^i \partial_i)
H_5][(\partial_t + v^i \partial_i)(\hat{H_1} \hat{H_K})] =
\left[(u^i-v^i) \hat{H_1^\lo} \partial_i H_5 \right] {\cal O}^{-1} \partial_a^2
\left[ (u^j-v^j) \hat{H_K^\lo} \pd_j H_5  \right], 
\end{equation}
where ${\cal O} = \tilde g^{\alpha \beta} \partial_\alpha \partial_\beta$. 
Now, convergence of the integral of the right hand side turns out to be
somewhat subtle when a homogeneous part is included, and depends upon
the detailed order in which certain limits are taken.  However, by 
treating the homogeneous part separately and realizing that it will not
contribute, we will avoid confusion. 

In the above form 
one can readily take the limit in which the sources
become distributions describing the desired branes,
\begin{eqnarray} 
\label{brlimit.eq}
c_{1,K}\tr_{1,K}^\dl & \rightarrow & 16\pi^2 \q_{1,K}^\dl  \delta^4(\vx
-\vx_5) \nonumber \\ 
c_{1,K}\tr_{1,K}^\lo & \rightarrow & 16 \pi^2 L^4 \q_{1,K}^\lo  \delta^4(\vx
-\vx_0) \delta^4(\vy -\vy_0) \nonumber \\
c_5\tr_5 & \rightarrow & 16 \pi^2 \qv  \delta^4(\vx -\vx_5). 
\end{eqnarray} 
Here $\vx_5$ is the position of the stack of 5-branes and the delocalized
1-branes and momentum, and $(\vx_0,\vy_0)$ is the position of the localized
stack of branes. $\q_{1,K}^\dl,\q_{1,K}^\lo$ are the charges of the
delocalized and the localized stacks of branes, respectively.  Note that
from the form of (\ref{sr.eq}) one can see that the details of this limit
are unimportant once the branes are localized on a scale much smaller than
the typical scale of variation of the functions $H_I^\lo$, $H_I^\dl$,
$H_5$.  Thus, for sufficiently large 4-torus and $r_5=\sqrt{\qv}$,
replacing the singular perfectly localized brane with a small cloud of
well-localized 1-brane charge and momentum charge yields identical results
in a regime in which supergravity is a valid description of the system.

Choosing an instantaneous coordinate system centered on the 5-brane, a
decomposition into modes along the torus shows that the
functions $H_5, H^\dl_1,H^\dl_K, H^\lo_1, H^\lo_K$ are given \cite{ms98} by 
\begin{eqnarray} 
\label{solution.eq}
H_5  &=& 1+ {\qv\over r^2} \nonumber \\
H_{1,K}^\dl  &=& {\q_{1,K}^\dl \over r^2} \nonumber \\
H^\lo_{1,K}(\vr,\vy;\vr_0,\vy_0) &=&  \sum_{l,\vq, m,n}
\fH_{1,K(l \vq)}(r;r_0)D^{\ast l}_{mn}(\psi_0,\theta_0,\phi_0)D^{l}_{mn}(\psi, \theta,\phi) \nonumber \\ 
&& \quad \times e^{i\vq \cdot (\vy-\vy_0)}
\end{eqnarray} 
where $(\psi, \theta,\phi)$ are the Euler angles on $S^3$, $\vec q ={2\pi
\vec n \over L}$, with $n\in {\mathbb Z}$, runs over the momentum lattice
of the torus, $D^l_{mn}(\psi, \theta,\phi)$ (including both integral and
half-odd integral $l$) are the rotation matrices which form a complete set
of functions on $S^3$ (see Appendix), and the radial functions
$\fH_{1,K(l\vq)}(r;r_0)$ are given by,
\begin{eqnarray} 
{\rm For} \qquad \vq=0, && \nonumber \\
\fH_{1,K( l 0)}(r;r_0) & = & { \q^\lo_{1,K}  \over (2l+1)} 
{r_0^{2l} \over r^{2l+2}} \qquad   r>r_0\nonumber \\
& = & { \q^\lo_{1,K}  \over (2l+1)} {r^{2l}\over  r_0^{2l+2}} \qquad
r<r_0  \label{deloc.eq} \\ 
{\rm and} \qquad |\vq| \neq 0 && \nonumber \\
\fH_{1, K( l \vq)}(r;r_0) & = & 2q \q^\lo_{1,K} {1\over rr_0} I_\mu (qr_0)K_\mu(qr) \qquad
r>r_0\nonumber \\
	& = & 2 q \q^\lo_{1,K} {1\over rr_0} K_\mu(qr_0) I_\mu(qr) \qquad
r<r_0 \label{loc.eq}, 
\end{eqnarray} 
where $\mu^2= 1+ 4l(l+1)+q^2\qv= 1+4l(l+1) + {4\pi^2 \over L^2}n^2\qv$. 
Note that the homogeneous ($\vec q=0$) modes
(\ref{deloc.eq}) satisfy the naive conservation equation:
\begin{equation}
\label{ConsEq}
\left( \frac{\partial}{\partial t} + v \cdot \nabla \right) \sum_l \fH_{1,K
(l,0)} = 0.
\end{equation}
In contrast, the inhomogeneous modes (\ref{loc.eq}) do not. 

This gives $H^\lo_{1,K}(0)={ \q^\lo_{1,K} \over r_0^2}$,
$H^\dl_{1,K}(\vx_0)={ \q^\dl_{1,K} \over r_0^2}$, $H_5(\vx_0)=1+ { \qv
\over r_0^2}$ so that (\ref{sr.eq}) simplifies to 
\begin{eqnarray}
\label{seff.eq}
S_{eff} & = {\pi \over G_5}& \int dt \{ -M + {1\over 2}[v^iv^i +v^av^a] \left( \qi^\lo + \qk^\lo 
\right) +
{1\over 2}u^iu^i \left( \qi^\dl + \qk^\dl + \qv \right) 
\nonumber \\ && +  {1
\over 2} (u^i -v^i)(u^i-v^i)
{{ \qv(\qi^\lo+ \qk^\lo) + \qi^\lo \qk^\dl +\qk^\lo \qi^\dl} \over {r_0^2}}
 + {{\qi^\lo \qk^\dl + \qi^\dl \qk^\lo} \over {2r_0^2}} 
v^a v^a\nonumber \\ && + 
 {1\over 2 } (u^i-v^i)(u^i-v^i) {{\qv [
{\qi^\lo\qk^\lo + 
 \qi^\lo \qk^\dl +\qk^\lo \qi^\dl} ]} \over {r_0^4}} \}
 \nonumber \\ && + {1\over 16 \pi  L^4 G_5} \int d^4xd^4y
\left[(u^i-v^i) \hat{H_1^\lo} \partial_i H_5 \right] {\cal O}^{-1} \partial_a^2
\left[ (u^j-v^j) \hat{H_K^\lo} \pd_j H_5\right] \nonumber \\
\end{eqnarray}  
where $M= \qv + \qi^\dl +\qk^\dl + \qi^\lo +\qk^\lo$ is the total
charge/mass.  

That is to say that the action is exactly the same as in the case \cite{gp99}
where both branes are delocalized, except for the inclusion of terms
involving $v^a$ and the addition of the last term,
\begin{equation} 
\cI= {1\over 16 \pi  L^4 G_5} \int d^4xd^4y
\left[(u^i-v^i) \hat{H_1^\lo} \partial_i H_5 \right] {\cal O}^{-1} \partial_a^2
\left[ (u^j-v^j) \hat{H_K^\lo} \pd_j H_5 \right]
\label{I.eq}
\end{equation} 
which remains to be evaluated.

\subsection{The Effective Action in the Near Horizon Limit}

It is difficult to obtain an analytic expression from the radial integral
in $\cI$, which involves products of three Bessel functions.  However, an
explicit result can be obtained in the limit $r_0 << {\sqrt \qv}$ in which
the localized branes are close to the 5-brane horizon.  In this case, the
Bessel functions of (\ref{loc.eq}) are approximated by powers of $r$.

An important fact is that since the $\vec q = 0$ modes (\ref{deloc.eq})
satisfy the conservation equation (\ref{ConsEq}), they  do not contribute to
(\ref{I.eq}) due to the factor of $\partial_a^2$ in that term.
This fact is true whether or not we have $r_0 \ll \sqrt{\qv}$.  A key
feature which becomes apparent here is that $\cI$ vanishes when either of
the localized charges are set to zero. For this special case, the
appearance of new moduli in the theory along the $y$ directions does not
influence the moduli metric in the transverse directions.  One expects that
this is related to the fact that setting one of the charges to zero doubles
the number of supersymmetries.

Explicitly,  we have
\begin{eqnarray} 
\cI &=& {\pi \over (16 \pi^2  L^4)^2 G_5 }\times  \qi\qk (u^i-v^i)(u^j-v^j)
\nonumber \\
&&\int d^4{\vec r} d^4 \vec y \Biggr[ \pd_i H_5(\vec r) \hat G(\vec
r,\vec y; \vec {r_0}, \vec {y_0}) \times \nonumber \\ 
&& \int d^4 \vec {r''} d^4 \vec {y''} \left[G(\vec
{r''},\vec {y''}; \vec r, \vec y) \pd_j H_5(\vec {r''}) \pd_a^2 \hat G(\vec
{r''},\vec {y''}; \vec {r_0}, \vec {y_0})\right]\Biggr], 
\end{eqnarray} 
where the Green's function $G(\vec r,\vec y; \vec r_0, \vec y_0)$ satisfies, 
\begin{equation} 
\cO G(\vec r,\vec y; \vec r_0, \vec y_0) 
= -16 \pi^2 L^4 \delta(\vec r-
\vec r_0) \delta (\vec y - \vec y_0), 
\end{equation} 
and $\hat{G}$ is the Green's function without it's homogeneous ($q=0$)
part.

Expanding in terms of the modes, 
\begin{eqnarray} 
\cI &=& {4 \pi \over (16 \pi^2  L^4)^2 G_5 }  \qi^\lo \qk^\lo \qv^2
(u^i-v^i)(u^j-v^j) \nonumber \\ 
&& \sum_{{q_1> 0, l_1,m_1,n_1} \atop {{q_2>0, l_2,m_2,n_2} \atop {q_3\ge 0, 
l_3,m_3,n_3}}}
q_1^2 D^{\ast l_1}_{m_1 n_1}(\Theta_0)\, D^{ l_2}_{m_2
n_2}(\Theta_0) \, e^{-i(\vec q_1 -\vec q_2)\cdot \vec y_0} \nonumber \\
&& \int dr \left[{\hat \fG_{(q_1l_1)}}(r;r_0) \int dr'' \fG_{(q_3 l_3)}(r'';r)
{\hat \fG_{(q_2l_2)}}(r_0;r'')\right] 
\nonumber \\ 
&& \int d\theta d\psi d\phi\, \left[ \sin\theta \, D^{l_1}_{m_1n_1}( \psi,
\theta, \phi)
\, D^{\ast l_3}_{m_3n_3}(\psi, \theta , \phi) \,b_i(\psi, \theta, \phi)\right]
\nonumber \\ 
&& \int d\theta'' d\psi'' d\phi'' \, \left[\sin\theta'' \,
D^{*l_2}_{m_2n_2}( \psi'', \theta'',\phi'') \, 
D^{l_3}_{m_3n_3}(\psi'', \theta'',  \phi'') \, b_j(\psi'', \theta'', \phi'')
\right] \nonumber\\
&& \int d^4y \left[e^{i(\vec q_1 -\vec q_3)\cdot \vec y}\right] \, \, \int
d^4y'' \left[e^{i(\vec q_1 -\vec q_3)\cdot \vec {y''}}\right]   
\end{eqnarray} 
where $\Theta_0$ represents the collective angular coordinates of the
stack of localized  1-branes, and we have used  
\begin{equation}
\label{defta.eq} 
\pd_iH_5 = -{2\qv \over r^3}{x_i \over r}= -{2\qv \over r^3}b_i(\theta,\psi,
\phi), 
\end{equation}
which defines $b_i$. 
%Note that the term proportional to $\delta(x)$ will not contribute to $\cI$.
%This can be seen from the fact that only the $q=0$ modes of $H^\lo_{1,K}$
%are nonzero at the 5-brane, while we have already found that such modes
%give vanishing contribution.

The radial integral can be evaluated in the near
horizon limit.  We need only consider the inhomogeneous ($\vec q \neq 0$)
modes (\ref{loc.eq}) which are approximated by the functions,
\begin{eqnarray} 
\label{nhl.eq}
\fG_{(ql)}(r;r_0)&\approx& {1 \over \mu} {r_0^{(\mu-1)}\over r^{\mu+1}} \qquad
r>r_0 \nonumber \\
& \approx&  {1 \over \mu}  {r^{\mu-1}\over r_0^{(\mu+1)}}
\qquad r<r_0.  
\end{eqnarray}
This gives us, 
\begin{equation} 
\label{I2.eq} 
\cI =  {4 \pi  \over (16 \pi^2)^2 G_5 } {\qi^\lo \qk^\lo \qv^2 \over
r_0^4}(u^i-v^i)(u^j-v^j)  \sum_{q > 0,l_1, l_2,l_3 }
q^2\zeta(\mu_1,\mu_2,\mu_3)  \times \fA_{ij}^{l_1l_2l_3}(\Theta_0) 
\end{equation} 
where, $\zeta(\mu_1,\mu_2,\mu_3)$ is given by
\begin{eqnarray}
\label{zeta.eq}
\zeta(\mu_1,\mu_2,\mu_3) &=& {4 \over \mu_1 \mu_2 \mu_3} {1 \over
(-2 + \mu_1 + \mu_2)(2 + \mu_1 + \mu_2)} \nonumber \\ && \times {
{(\mu_1 + \mu_2)(\mu_1 + \mu_3)(\mu_2 + \mu_3)(\mu_1 + \mu_2 + \mu_3)
+ \mu_3(\mu_1 + \mu_2)  + 4 \mu_1 \mu_2 } \over (-1 + \mu_1 +
                  \mu_3) (1 +
                  \mu_1 + \mu_3)(-1 + \mu_2 + \mu_3)(1 + \mu_2 +
                  \mu_3)} \nonumber, \\
\end{eqnarray}
and  
\begin{eqnarray} 
\label{defA.eq}
\fA_{ij}^{l_1l_2l_3}(\Theta_0) &= & \sum_{m_1,m_2,m_3, n_1,n_2,n_3} \, D^{\ast 
l_1}_{m_1 n_1} (\Theta_0) \, D^{l_2}_{m_2 n_2} (\Theta_0) \nonumber \\ 
&& \int d\theta d\psi d\phi \quad \left[\sin \theta \,
D^{l_1}_{m_1n_1}(\psi,\theta,\phi)\, D^{\ast
l_3}_{m_3n_3}(\psi,\theta,\phi)\, b_{i}(\psi,\theta,\phi)\right]
\nonumber \\  
 && \int  d\theta'' d\psi'' d\phi'' \quad \left[ \sin \theta''\,
D^{\ast l_2}_{m_2n_2}(\psi'',\theta'',\phi'') \,
D^{l_3}_{m_3n_3}(\psi'',\theta'',\phi'') \,
b_{j}(\psi'',\theta'',\phi'')\right]. 
\nonumber \\     
\end{eqnarray} 
Note that $\zeta$ is symmetric in $\mu_1,\mu_2.$
Each component of $b_i(\psi,\theta,\phi)$ is a sum over two rotation
matrices, so that the angular integration is easily performed using an
identity (\ref{3d.eq}) from the Appendix.

In order to evaluate $\fA_{ij}(\Theta_0)$, we note that there are two
determining vectors in the transverse directions: the relative transverse
velocity of the stack of localized 1-branes with respect to the 5-brane
$(v^i-u^i)$, and the transverse separation vector $\vec{x}_0$ between the 5
and 1-brane.  We may then choose coordinates so that both sets of branes
lie in the 1-2 plane.

	In order to ease the calculation, we consider the instantaneous
frame in which the 5-brane is at the origin, and the stack of localized
1-branes is on the 1-axis. Symmetry about the 1-axis dictates that the
off-diagonal part of $\fA_{ij}$ is zero, and that
$\fA_{22}=\fA_{33}=\fA_{44}$.  On the 1-axis, $\theta_0,\psi_0,\phi_0=0$,
and thus $D^{\ast l}_{mn}(0,0,0)=\delta_{mn}$ reduces the number of
summations. This simplifies the angular part and we find (see Appendix),
\begin{eqnarray} 
\label{FH2.eq}
(u^i-v^i)(u^j-v^j)\fA^{l_1l_2l_3}_{ij}(\Theta_0) &=& {(8\pi^2)^2 \over 2}
\delta({1\over 2},l_1,l_3)\delta({1\over 2},l_2,l_3) \nonumber \\& & 
[(F+ H)^{l_1l_2l_3}(v^r-u^r)^2 -
(F-H)^{l_1l_2l_3}r_0^2(v^\phi-u^\phi)^2]\nonumber \\ 
\end{eqnarray} 
where $F^{l_1l_2l_3}$ and $H^{l_1l_2l_3}$ are given by (\ref{FH.eq}) in the
Appendix and 
$\delta(j,k,l)$ is the triangle condition,  
\begin{eqnarray}
\label{triangle.eq}
\delta(j,k,l) &=& 1 \ {\rm for} \
j+k \geq l \geq |j-k|, \ \rm{and} \ j+k-l \ \rm{an \ integer} \nonumber \\ 
\delta(j,k,l) &=& 0, \ {\rm otherwise}.  
\end{eqnarray}

Thus, 
\begin{equation} 
\cI =  {\pi \over 2 G_5}{\qi^\lo\qk^\lo \qv  \over r_0^{4}} [(\f_0
+\h_0)(v^r-u^r)^2 -(\f_0 -\h_0)r_0^2(v^\phi-u^\phi)^2], 
\end{equation} 
where
\begin{eqnarray} 
 \label{fdef.eq}
\f_0 &=& \sum_{q> 0,l_1,l_2,l_3} \qv q^2 \zeta(\mu_1,\mu_2,\mu_3) F^{l_1l_2l_3}
\delta({1\over 2},l_1,l_3)\delta({1\over 2}, l_2,l_3)\nonumber \\
\h_0&=& \sum_{q> 0,l_1,l_2,l_3} \qv q^2 \zeta(\mu_1,\mu_2,\mu_3) H^{l_1l_2l_3}\delta({1\over 2},l_1,l_3)\delta({1\over 2}, l_2,l_3). 
\end{eqnarray} 

As noted above, the $q=0$ (`homogeneous') modes do not contribute to
(\ref{fdef.eq}).  Note that for large $r_5/L$ the $n^2 = q^2 L^2/4\pi^2
>0$ contributions are highly suppressed by the correspondingly large values
of $\mu$ even for the lowest term $n=1$.  Thus, both $\f_0$ and $\h_0$
vanish in the limit of large $r_5/L$.  On the other hand, for small
$r_5/L$, our $\mu$ (and therefore the quantity to be summed in
(\ref{fdef.eq})) depends only weakly on the integer $n^2 = L^2 q^2/4\pi^2$
and many terms contribute with equal weight.  Thus, $\f_0$ and $\h_0$ are
correspondingly large in this limit.  In section \ref{disc} we will
discuss in more detail the physically appropriate way to take the $L
\rightarrow \infty$ limit.  However, for now we simply note that, since $n$
appears in $\mu$ only through $\frac{\qv}{L^2}4 \pi^2 n^2$, the growth of
$\f_0,\h_0$ for large $L$ is of the form $\frac{\qv}{L^2} \int_{n^2 \le
\frac{L^2}{4 \pi^2 \qv}} n^2 d^4n \sim \left( \frac{L^2}{\qv} \right)^2.$ 

Thus, the effective Lagrangian we obtain to leading order in the near
horizon limit is, 
\begin{eqnarray}
\label{leff.eq}
L_{eff} & \approx & {1\over 2 r_0^2} \left(\qi^\lo\qk^\dl + \qi^\dl\qk^\lo
 \right) v^av^a 
 \nonumber \\ 
&+& {1 \over 2r_0^4}\qv (\qi^\lo\qk^\dl  +  \qi^\dl\qk^\lo +
 \qi^\lo\qk^\lo(1+(\f_0+\h_0) ) (v^r-u^r)^2 \nonumber \\ &+& {1 \over
 2r_0^4}\qv 
(\qi^\lo\qk^\dl  +  \qi^\dl\qk^\lo + 
 \qi^\lo\qk^\lo(1 - (\f_0 -\h_0))r_0^2(v^\phi -u^\phi)^2. \nonumber \\
\end{eqnarray} 
Note that the zeroth order velocity contribution here is  a constant
potential equal to the total mass. Hence the dynamics of the system is
determined entirely by the geodesics on the moduli space metric.

\subsection{The Moduli Space}
\label{ModSpace}

It is useful to cast the effective action in the center of mass
coordinates 
in which the relative velocity is  $\omega^\alpha= v^\alpha-u^\alpha$. 
The moduli space metric to leading order in the near horizon limit is
therefore, 
\begin{eqnarray} 
\label{mmm.eq}
ds^2 &\approx & {\qv \over r_0^4} (\qi^\lo\qk^\dl  + \qi^\dl\qk^\lo
+\qi^\lo\qk^\lo (1+[\f_0+\h_0])) dr_0^2 \nonumber \\  
& +& {\qv \over r_0^4} (\qi^\lo\qk^\dl  + \qi^\dl\qk^\lo +
\qi^\lo\qk^\lo (1-[\f_0-\h_0])) r_0^2d\Omega_3^2  \nonumber \\ 
& + & {(\qi^\lo\qk^\dl + \qi^\dl\qk^\lo) \over r_0^2}
dy^ady^a,  
\end{eqnarray} 
where $d\Omega_3^2$ is the metric on the unit 3-sphere. Relevant 
quantities are the total mass 
$M = \qv + \qi^\dl+\qk^\dl + \qi^\lo +\qk^\lo$, the
reduced mass $\mmu = {( \qi^\lo + \qk^\lo )( \qv +
\qi^\dl +\qk^\dl) \over M}$, 
and the center of mass velocity $V^\alpha = { ( \qi^\lo + \qk^\lo)v^\alpha
+ (\qv + \qi^\dl +\qk^\dl) u^\alpha \over M}$.
However, the center of mass terms do not appear explicitly in
(\ref{mmm.eq}) since the center of mass part of the metric is constant in
these coordinates and we have restricted our analysis to the leading order
$r_0^{-4}$ contribution.  The metric (\ref{mmm.eq}) has a warped product
structure, with the transverse radial direction warping the metric in the
internal directions.  At first sight, it may appear odd that the $dy^a
dy^a$ term does not depend on $\qv$.  Recall however, that a fundamental
string (without longitudinal momentum) should respond to the string metric,
at least in the test string approximation.  Thus, the above result might be
expected from the fact that, in the string frame, the metric for a
Neveu-Schwarz fivebrane is simply $dy^ady^a$ in the torus directions.

An important question which arises at this stage is whether the metric has
singularities. Notice that if $\f_0 -\h_0>0$, then, by tuning the values of
the charges one could make the coefficient of $d \Omega_3^2$ in
(\ref{mmm.eq}) negative\footnote{We thank Andy Strominger for pointing this
out.}.  Since (\ref{mmm.eq}) describes the leading near-horizon behavior,
such an effect could not be compensated by the neglected terms.  However,
as we demonstrate in the Appendix, this is not the case: namely, $\f_0-\h_0
<0$.

We have only calculated the metric in the near horizon limit $r_0
\rightarrow 0$.  There, a change of coordinates $r_0 = \rho^{-1}$ 
illustrates the fact that, as in the delocalized case,  
 small $r_0$ is really a large asymptotic region of the relative
motion moduli space.  The difference between the present case and the
delocalized case is that the coefficients of the $d\rho^2$ and $\rho^2
d\Omega_3^2$  terms do not agree.  As a result, our transverse moduli space
is not quite asymptotically flat in this region.  Nonetheless, curvature
scalars do go to zero for small $r_0$.  
%and calculations in this region show
%that there must be some critical impact parameter $b_c$ such that black
%holes which approach each other closer than $b_c$ necessarily coalesce.

More specifically, let us consider the special case when the $T^4$
velocity is zero. Again suppose that the motion takes place in
some plane so that only one angle $\phi$ on the 3-sphere is relevant.
We first rewrite  the effective action in terms of the
parameters $\chi$ and $\xi$ as,  
\begin{equation} 
L_{eff}={1 \over 2 r_0^4}\chi {\dot r}^2 + {1\over 2 r_0^2}\xi {\dot \phi}^2.  
\end{equation} 
%their asymptotic forms, ${1 \over 2} \mmu v_\infty^2$ and $\mmu b
%v_\infty$, respectively, where
%\begin{equation} 
%\mmu= {( \qi^\lo + \qk^\lo )( \qv + \qi^\dl +\qk^\dl) \over {\qv + \qi^\dl
%+\qk^\dl + \qi^\lo +\qk^\lo}}, 
%\end{equation} 
%is the reduced mass for the system, $v_\infty$ is the asymptotic transverse
%velocity and $b$ is the impact parameter. 
If the conserved energy and momentum for this system are ${\cal E}$ and
$\cal L$ respectively, the effective radial motion in
the near horizon region is governed by the equation,
\begin{equation} 
{\dot r_0}^2 + {2 {\cal E}\over \chi}[{{\cal L}^2 \over
2 {\cal E} \xi}r_0^6 -r_0^4] =0.  
\end{equation} 
The classically accessible regions are those for which the effective
potential $U_{eff} = {2 {\cal E}\over \chi}[{{\cal L}^2 \over 2 {\cal E}
\xi}r_0^6 -r_0^4] \leq 0$ and the turning point for the radial motion occurs
when
\begin{equation} 
\label{luco.eq}
r_t= \sqrt{{2 {\cal E} \xi\over {\cal L}^2 }},     
\end{equation}  
and at $r_t=0$. Thus, $U_{eff}\leq 0$ only for $r<r_t$, so that the branes
are confined to this region.    
$U_{eff}$, moreover, has a minima at  
\begin{equation} 
r_m=\sqrt{2 \over 3} r_t. 
\end{equation} 
>From (\ref{luco.eq}), it would appear that there is a turning point for the
motion for any value of the angular momentum, thus at variance with the
delocalized case where the branes can sometimes escape to infinity.
%% for which there is a critical impact parameter beyond
%% which widely separated branes only coalesce. 
However, it must be noted that
(\ref{luco.eq}) is valid only in the near horizon region: for
sufficiently small angular momentum, $r_t$ lies outside the
near horizon region.

On physical grounds we expect minimal interaction of the objects
at large distances, so that at large $r$ the metric should be asymptotically
flat as in \cite{km}.  Thus, black hole scattering should have
the familiar qualitative behavior of \cite{km} with a critical impact
parameter, 
depending on the various charges, ${\cal E}$, and ${\cal L}$, which
separates coalescing orbits from orbits for which the branes
escape to relative infinity.

As we noted before, $\cI$ vanishes when the localized momentum $\qk^{\lo}$
is set to zero. 
In this case we need not limit our analysis to the near 
horizon region and the effective action (\ref{seff.eq}) yields the
moduli metric,
\begin{eqnarray} 
\label{qkzero.eq}
ds^2 &= & \Biggr(\mmu + {[\qv \qi^\lo
+ \qi^\lo\qk^\dl ] \over  
r_0^2} + {{\qv \qi^\lo\qk^\dl } \over r_0^4} 
\Biggr) dr_0^2 \nonumber \\ 
& +&   \Biggr(\mmu  +{[\qv \qi^\lo 
+ \qi^\lo\qk^\dl ] \over  
r_0^2}  +  {{\qv \qi^\lo\qk^\dl  } \over r_0^4} 
\Biggr) r_0^2d\phi_0^2  \nonumber \\ & + &
\Biggr(\mmu + {\qi^\lo\qk^\dl  \over r_0^2} \Biggr)
dy^ady^a.  
\end{eqnarray}  
The transverse part of this metric coincides with the results of the black
hole calculation \cite{km,gp99} for this set of charges.

In particular, when $\qk^{\dl}=0$, this moduli metric
reduces to the particularly simple form,   
\begin{equation} 
\label{nomom}
ds^2 =  [\mmu + {\qv \qi^\lo \over r_0^2}] [dr_0^2 + r_0^2 d\Omega_3^2] + \mmu
dy^ady^a,  
\end{equation} 
thus reproducing the probe calculation of \cite{dps}.

With non-zero localized momenta, however, even to leading order in the near
horizon limit, (\ref{mmm.eq}) differs from the black hole moduli space
calculations of \cite{km,gp99} in detail, even though some gross features
are preserved.  In particular, the transverse moduli space metric for
relative motion is no longer conformally related to the standard
Euclidean space metric
given by the isotropic coordinates.
We also note that the coefficients $\f_0$ and $\h_0$ are now
functions of the ratio $r_5/L$. This remains true even when the extra
charges $\qi^\dl, \qk^\dl$ on the 5-brane are set to zero, and hence can be
seen as a generic feature of brane localization.

\section{Discussion} 
\label{disc}

Our results describe the moduli space for a stack of localized 1-branes
interacting with a stack of 5-branes.  Both branes are allowed to
carry momentum in the direction along the 1-brane, and the 5-brane is also
allowed to carry 1-brane charge.  All charges on the 5-brane are necessarily
delocalized along the 5-brane.

If the localized branes are replaced with a system in which either the
one-brane charge or the momentum charge is delocalized, the structure of
the moduli space simplifies greatly and reduces to previously known forms
(e.g. \cite{km,gp99,dps}). When the momentum vanishes, the fact the the
moduli space is independent of whether the 1-brane charge is localized
might be expected from the (4,4) nonrenormalization theorem described in
\cite{dps}.  That the simple form persists in the presence of delocalized
momentum charge is interesting, since momentum charge breaks the same
supersymmetries whether or not it is localized.

Another interesting point is that localization affects the structure of the
moduli space even in the near 5-brane limit.  Recall that  when
a one-brane approaches a 5-brane, there is a sense \cite{ms98} in which
it `spontaneously' delocalizes.  Because of this, one might have
expected the moduli space for localized 1-branes to go over to that
of delocalized one-branes in the near 5-brane limit.  However, this is
not the case.  The reason for this is that (see \cite{ms98})
the one-branes only appear to spontaneously delocalize from the viewpoint
of an observer far from the one-brane.  When one examines the solutions
in the immediate vicinity of the one-brane, it is clear that the one-brane
is in fact localized.  

	Thus, the effective action is sensitive to the region
near the one-brane and thus to the localization.  It was shown in
\cite{pm99} how the spontaneous delocalization is described in the dual
field theory, but it is less clear which field theory observable would
encode the fact that the one-brane is localized as viewed by a nearby
observer. As a result, it would be  interesting to discover how our moduli
space metric can be understood from the dual field theory description. 

In this near 5-brane limit, we were able to study the structure of the
moduli space for this system in some detail. Our results differ from those
of previously known, less localized cases \cite{km,gp99,dps}, through a
modification of the three-charge term\footnote{Since the
coefficient of this term now involves a complicated function of
$\qv/L^2$, it is not clear that the terminology ``three-charge term''
is strictly speaking appropriate.  Nevertheless, it is a convenient
way to refer to this term.}.  Although our setup is somewhat
different, it is interesting to note that a three-charge term is
responsible for the puzzle described in \cite{dps}.  Thus, such terms may
warrant further consideration in the future.

Although the three-charge term is modified relative to the less localized
case, scattering in the localized moduli space must exhibit the same
qualitative behavior  as in \cite{fe,km,dps}. 
%%We find that in 
%%the near horizon region, the effective motion is bounded by a maximal
%%radius so that the branes are confined.  
%and thus doesn't represent scattering of widely separated branes.     
>From the analysis of \cite{fe,km,dps}, we know that for the
delocalized case there is a critical impact parameter beyond which widely
separated branes always coalesce. In the near horizon limit, however, we
see that this critical impact parameter cannot be calculated.

%there are two asymptotic regions, one at
%$r\rightarrow \infty$ and the other at $r \rightarrow 0$, and there is a
%critical impact parameter within which the branes coalesce. 

As the coefficients $\f_0$ and $\h_0$ of our three-charge terms are
complicated functions of the ratio $r_5/L$, it is enlightening to discuss
their behavior in various limits.  We have seen that they are large ($\sim
L^4/\qv^2$) for $r_5 \ll L$ with $\qi,\qk$ fixed.  For $r_5 \gg L$ and
$\qi,\qk$ fixed, the behaviors of $\f_0$ and $\h_0$ are controlled by the
behavior of $\zeta$ for the lowest modes with $n^2=1$.  Since, $\zeta$
scales like $\mu^{-5}$ for large $\mu$, we see that $\f_0,\h_0 \sim
\left(\frac{L^2}{\qv} \right)^{\frac{3}{2}}$ in this limit.  Such scalings
would correspond to, for example, changing the charge on the fivebrane
while holding all other parameters fixed.

Changing the size of the torus, however, is not naturally described by such
a limit.  Presumably, it is more appropriate to change the size of the
torus holding fixed the {\it ten}-dimensional 
parameters.  This is equivalent to holding fixed our 9-dimensional
parameters as there is no need to rescale the size $L_z$ of the remaining 
circle.  As the dimensions of one-brane and momentum
charge in ten dimensions is naturally $(length)^6$, to hold fixed the
ten-dimensional parameters we should scale each of $\qi,\qk$ as $L^{-4}.$
We should also include the overall factor of $1/G_5$ in the effective
action, and holding fixed the ten-dimensional Newton's constant will cause
$G_5$ to also scale as $L^4.$ Taken together with the divergence of
$\f_0,\h_0$, we see that our term makes a finite non-zero contribution in
this large $L$ limit although the usual three-charge term becomes
vanishingly small.

On the other hand, for a small torus with ten-dimensional parameters
held fixed, our new terms scale as $L^{-1}$.
While we see that these new terms do become large in this limit, the
standard three-charge term in fact scales as $L^{-4}$, so that our 
modification becomes irrelevant.

	It would be of interest to understand our moduli space metric as
the target space of a supersymmetric sigma model in the spirit of
\cite{mist,gps97,gp98}.  Although the effective theory we consider includes
extended objects, freezing the $T^4$ moduli reduces it to one of
point particles so that the relevant sigma model will be 1-dimensional as
in \cite{mist,gps97,gp98}. Moreover, in \cite{gp00} a general moduli
potential
\begin{equation}
\label{modpot}
\int d^9x  H_1 H_K H_5,
\end{equation}
where $d^9x$ is the Euclidean measure associated with isotropic coordinates
$x$, was proposed for a large class of 3-charge brane solutions preserving
4 supercharges. The localization of our charges means that our solution
falls outside the class of solutions considered there, but nevertheless it
does preserve four supercharges.  It is therefore of interest to know how
their proposed scheme may be extended to include the localized case. A
short calculation shows that a naive attempt to use (\ref{modpot}) directly
in our context would predict that, in the usual isotropic coordinates, the
transverse part of the moduli space metric (\ref{mmm.eq}) for single brane
scattering to be simply a conformal factor multiplied by the standard
Euclidean metric.  As discussed in section \ref{ModSpace}, this is not the
case \footnote {Nonetheless, the spherical symmetry of our two-body
transverse relative moduli space means that it is conformally flat in
different coordinates. This observation allows one to construct a
moduli potential, showing that our two-body moduli space is appropriately
supersymmetric.}.

In the introduction, a possible connection was mentioned to the work of
\cite{mist,mpsv00,psv00} which endeavors to associate internal states of
black holes with the multi-black hole moduli space.  As in their work for
the delocalized case, we find an asymptotic region of the moduli space when the
branes are nearly coincident.  Thus, as one would expect, the moduli space
for localized branes also has a continuum of low energy states.  In the
black hole case, a superconformal structure was discovered which allowed a
new choice of Hamiltonian with a discrete spectrum and finite density of
states.  It would be interesting to know if such a structure arises in this
case as well, though we leave this as an open question for the moment.
Taking into account the properties of localized branes may lead
to further developments for this program.

\section{Acknowledgments}
We would like to thank Sumit Das, Bernard Kay, Jeremy Michelson, Ashoke
Sen, and David Tong for helpful discussions.  We also thank Andy Strominger
and George Papadopoulos for pointing out errors in a previous draft and we
thank Jan Gutowski for his patient discussion of calculations associated
with the Chern-Simons term.  S. Surya was supported by a Visiting
Fellowship from the Tata Institute of Fundamental Research.  D. Marolf is
an Alfred P. Sloan Research Fellow and was supported in part by funds from
Syracuse University and from NSF grant PHY-9722362.

\section{Appendix }
\label{appendix2.sec}

We use the Euler angles on the 3-sphere, $(\psi,\theta,\phi)$, where $0
\leq\theta \leq  \pi$, $0\leq \phi,\psi \leq 2\pi$.   The
transformation to Cartesian coordinates is,  
\begin{eqnarray} 
x_1=r \cos {\theta \over 2} \cos{ \phi +\psi \over 2} \nonumber \\
x_2=r \cos {\theta \over 2} \sin{ \phi +\psi \over 2} \nonumber \\
x_3=r \sin {\theta \over 2} \cos{ \phi -\psi \over 2} \nonumber \\
x_4=r \sin {\theta \over 2} \cos{ \phi -\psi \over 2}. 
\end{eqnarray} 
To calculate $\fA_{ij}$, in the 1-2 plane,  we use the following:
\begin{eqnarray} 
b_1 &=& {1\over 2} [D^{{1\over2}}_{{1\over 2} {1\over
2}}+D^{{1\over2}}_{-{1\over 2} -{1\over 2}}] \nonumber \\
b_2 &=& {1\over 2 i} [D^{{1\over2}}_{{1\over 2} {1\over
2}} - D^{{1\over2}}_{-{1\over 2} -{1\over 2}}] \nonumber \\
\end{eqnarray} 
where $b_i(\psi,\theta,\phi)$ is given by (\ref{defta.eq}).  The following
identity,  from \cite{rose,bsat}, can be used to perform the angular
integrals in (\ref{defA.eq}),
\begin{equation} 
\label{3d.eq}
\int  d\theta d\psi d \phi \quad \left[ sin \theta \, D^l_{mn}\,  
	D^{l_1}_{m_1n_1} \, D^{l_2}_{m_2n_2}\right] = 8\pi^2  \times \left(\begin{array}{rrr} 
	l &  l_1 & l_2 \cr
	m &  m_1 & m_2 \end{array} \right) \times 
\left(\begin{array}{rrr} 
	l &  l_1 & l_2 \cr
	n  &  n_1 & n_2 \end{array} \right). 
\end{equation}

The result is that
$F^{l_1l_2l_3}$ and $H^{l_1l_2l_3}$ in (\ref{FH2.eq}) are given by, 
\begin{eqnarray} 
\label{FH.eq}
F^{l_1l_2l_3}&\equiv&Q^{l_1l_2l_3}\sum_{m_1} {(l_2+m_1+1)! (l_1-m_1)! \over
(l_2-m_1-1)! (l_1+m_1)!} \nonumber \\ 
H^{l_1l_2l_3}&\equiv&Q^{l_1l_2l_3}\sum_{m_1} {(l_2+m_1)! (l_1+m_1)! \over
(l_2-m_1)! (l_1-m_1)!}\Biggr[{(l_3 -m_1 +{1 \over 2})! \over  (l_3 +m_1 - {1
\over 2})!}\Biggr]^2,  \nonumber \\  
\end{eqnarray} 
where, 
\begin{equation} 
Q^{l_1l_2l_3}= {(-{1\over 2}+l_1+l_3)!(-{1\over 2}+l_2+l_3)! \over 
(l_1+l_3 +{3\over 2})! (l_2+l_3 +{3\over 2})!({1\over 2}+l_1 - l_3)!
({1\over 2}+l_2 - l_3)! ({1\over 2}+l_3 - l_1)! 
({1\over 2}+l_3 - l_2)! } 
\end{equation}   
and we have used the Wigner closed expression for the
Clebsch Gordon coefficients (see \cite{rose} chapter 3, for example).

In order to show that $\f_0 -\h_0 \leq 0$, we use the fact that the
triangle conditions, $\delta({1\over 2},l_1,l_3)$ and $\delta({1\over
2},l_2,l_3)$ impose rather severe restrictions on the sums over
$l_1,l_2,l_3$. This allows us to restrict to the four possible
cases for each term,
\begin{eqnarray} 
{\rm Case \, 1}: \,\, &  l_3 = l_1 -{1\over 2} ;\, \, \, & l_2=l_1 \nonumber \\ 
{\rm Case \, 2}: \, \,&  l_3 = l_1 +{1\over 2} ;\, \,\,& l_2=l_1 \nonumber \\ 
{\rm Case \, 3}: \, \,&  l_3 = l_1 +{1\over 2} ;\,\,\,& l_2=l_1+1\nonumber \\  
{\rm Case \, 4}: \, \, &  l_3 = l_1 -{1\over 2} ;\,\,\, &
l_2=l_1-1\nonumber \\  
\end{eqnarray} 
This helps us evaluate, 
\begin{eqnarray} 
\label{cases}
{\rm Case \,1} &:(H-F)^{J+1,J+1,J+{1\over2}}&={1 \over 12} {(2J+1) \over
(2J+3)(J+1)} 
\nonumber \\ 
{\rm Case \,2}  &: (H-F)^{J,J,J+{1\over2}}&={1 \over 12} {(2J+3) \over
(2J+1)(J+1)} 
\nonumber \\  
{\rm Case \,3} &: (H-F)^{J,J+1,J+{1\over 2}}&=-{1 \over 12} {1 \over (J+1)}
\nonumber \\  
{\rm Case \, 4} &: (H-F)^{J+1,J,J+{1\over2}} &=-{1 \over 12} {1 \over
(J+1)}, \nonumber \\  
\end{eqnarray} 
where we have put $l_3=J+{1\over 2}$ for each case. One may
check that, the summation
over $l_1,l_2,l_3$ reduces to a sum over $J$ which takes on integer and
half odd-integer
values, $J \geq 0$.  The corresponding form of
$\zeta(\mu_1,\mu_2,\mu_3)$ for each case $s\in \{1,2,3,4\}$, is a clumsy
expression, which we denote as $\zeta_s$. 

Now, it suffices to show that $\f_0-\h_0 <0$ term by term in the sum over 
$q$. 
Define $ c = 1+q^2\qv$, and 
\begin{equation} 
\U_s(c)= (c-1)\sum_J \zeta_s (H-F)^s,  
\end{equation} 
where 
\begin{equation}
\h_0-\f_0=  \sum_{q >0} \sum_s \U_s(c). 
\end{equation}
 
In order to test for positivity, we consider a truncated summation up to
$J=100$ for each case (\ref{cases}), which we refer to as $\wU_s(c)$. We
then plot the various $\wU_s(c)$'s in Fig(\ref{1234.fig}), as functions of
$c$. Note that $c >1 $ for the inhomogeneous modes we are considering, and
that $c=1$ doesn't contribute. Finally, plotting $\sum_s\wU_s(c)$, in
Fig(\ref{total.fig}), we find a distinctively positive function.  We have
also found that the result given by truncating the series at $J=5$ to be
essentially the same as that shown below, so that we believe the numerical
results to be accurate and the sum to converge rapidly. Figs (\ref{j1.fig},
\ref{j2.fig}, \ref{j34.fig}) demonstrate this convergence.

Thus, we conclude that term by term
in $q$, $\f_0 -\h_0 <0$.

%%%%%%%%%%%%%%%%%%%%%%%%%%%%%%%%%%%%
\EPSFIGURE{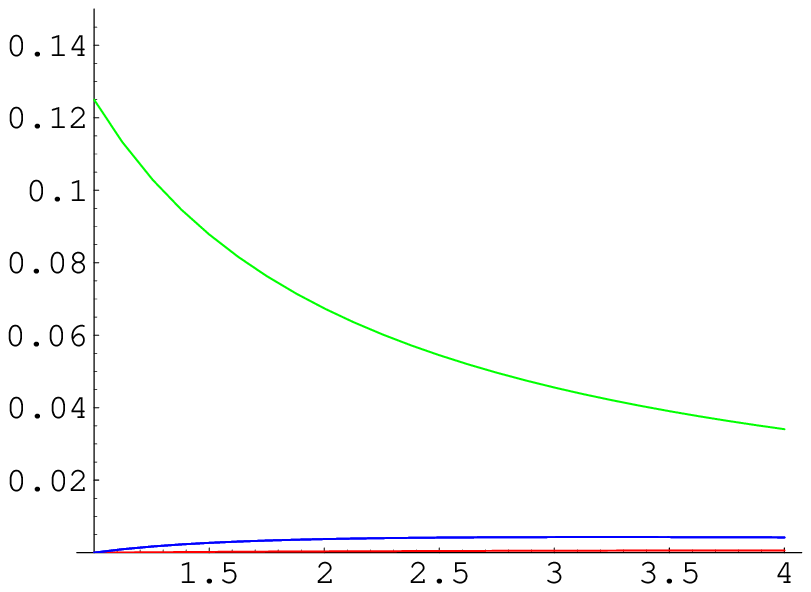,width=10cm}{$c$ is on the horizontal-axis and 
the $\wU_s$ are on the vertical-axis. $\wU_1$ is
represented by the red line, $\wU_2$ by the green line and $-\wU_{3,4}$ by
the blue line}\label{1234.fig}
%%%%%%%%%%%%%%%%%%%%%%%%%%%%%%%%%%%%%

%\begin{figure}[ht]
%\centering
%\resizebox{3.5in}{!}{\includegraphics{Cases1234.eps}}
%\caption{$c$ is on the horizontal-axis and 
%the $\wU_s$ are on the vertical-axis. $\wU_1$ is
%represented by the red line, $\wU_2$ by the green line and $-\wU_{3,4}$ by
%the blue line}\label{1234.fig}
%\end{figure}

%%%%%%%%%%%%%%%%%%%%%%%%%%%%%%%%%%%%
\EPSFIGURE{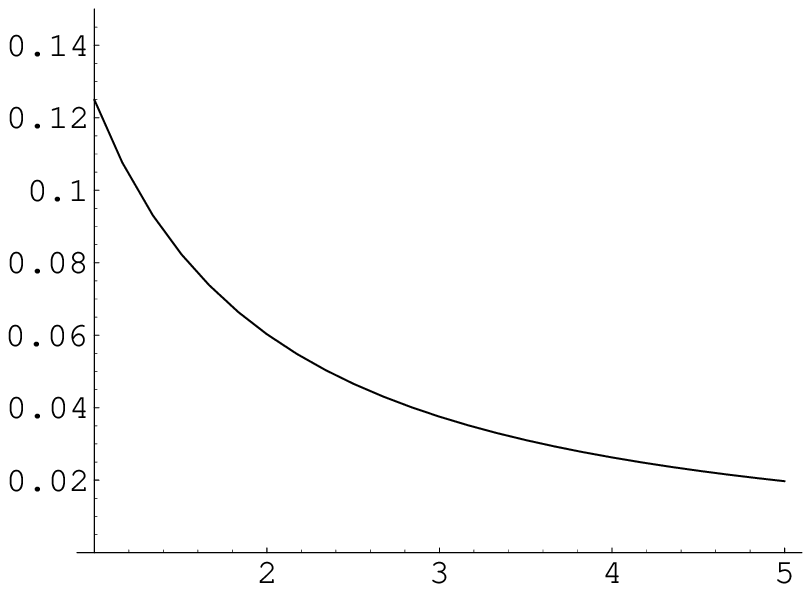,width=10cm}{
$c$ is on the horizontal-axis and $\sum_s\wU_s(c)$ is on the
vertical-axis}\label{total.fig} 
%%%%%%%%%%%%%%%%%%%%%%%%%%%%%%%%%%%%%

%\begin{figure}[ht]
%\centering
%\resizebox{3.5in}{!}{\includegraphics{Total.eps}}
%\caption{$c$ is on the horizontal-axis and $\sum_s\wU_s(c)$ is on the
%vertical-axis}\label{total.fig} 
%\end{figure}

%%%%%%%%%%%%%%%%%%%%%%%%%%%%%%%%%%%%
\EPSFIGURE{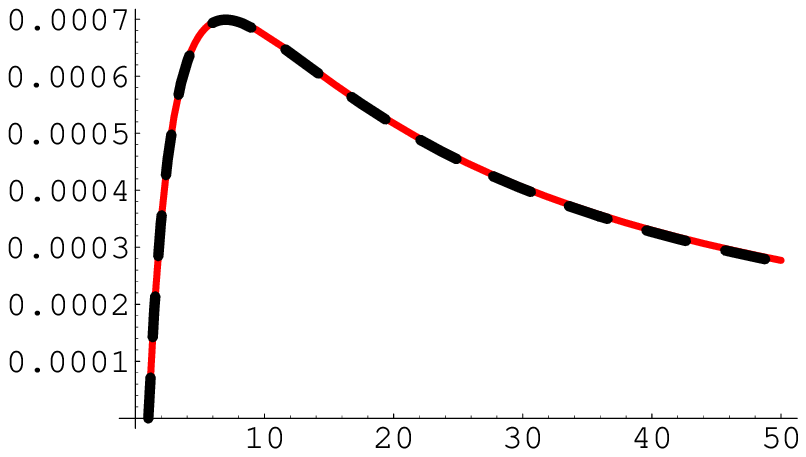,width=10cm}{
The red line is the sum $\U_1$ upto $J=100$ and the dashed black
line upto $J=5$. $c$ is on the horizontal-axis.}\label{j1.fig} 
%%%%%%%%%%%%%%%%%%%%%%%%%%%%%%%%%%%%%

%\begin{figure}[ht]
%\centering
%\resizebox{3.5in}{!}{\includegraphics{100-1.eps}}
%\caption{ The red line is the sum $\U_1$ upto $J=5$ and the dashed black
%line upto $J=100$. $c$ is on the horizontal-axis.}\label{j1.fig} 
%\end{figure}

%%%%%%%%%%%%%%%%%%%%%%%%%%%%%%%%%%%%
\EPSFIGURE{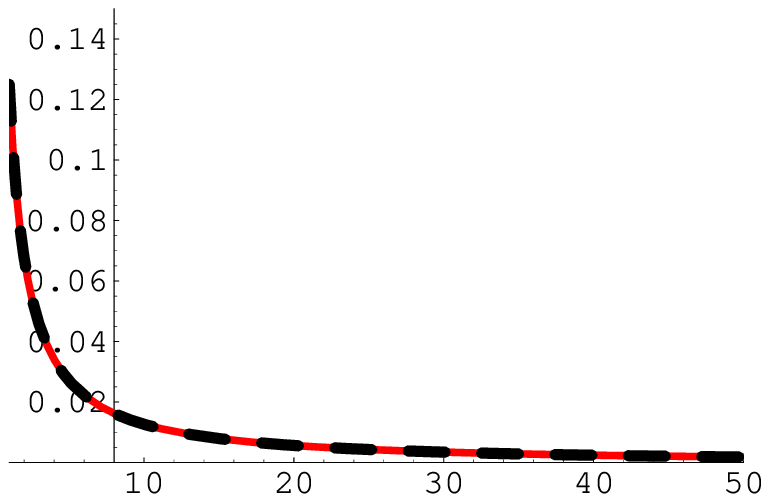,width=10cm}{
The red line is the sum $\U_2$ upto $J=100$ and the dashed black
line upto $J=5$. $c$ is on the horizontal-axis.}\label{j2.fig} 
%%%%%%%%%%%%%%%%%%%%%%%%%%%%%%%%%%%%%

%\begin{figure}[ht]
%\centering
%\resizebox{3.5in}{!}{\includegraphics{100-2.eps}}
%\caption{ The red line is the sum $\U_2$ upto $J=5$ and the dashed black
%line upto $J=100$. $c$ is on the horizontal-axis.}\label{j2.fig} 
%\end{figure}

%%%%%%%%%%%%%%%%%%%%%%%%%%%%%%%%%%%%
\EPSFIGURE{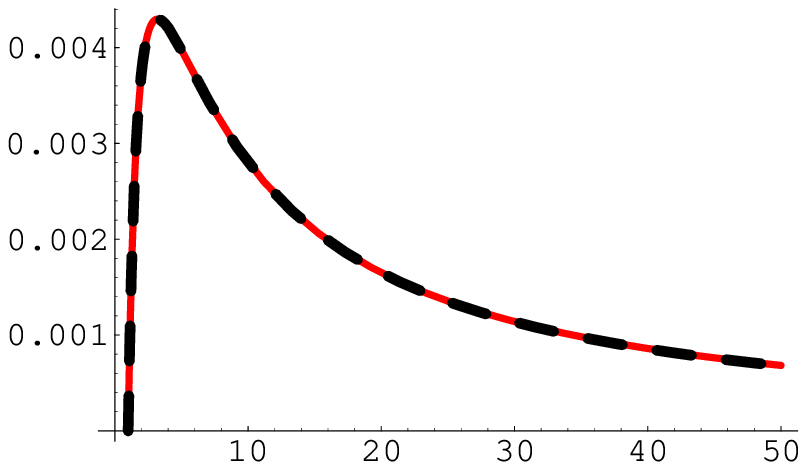,width=10cm}{
The red line is the sum $\U_{3,4}$ upto $J=100$ and the dashed black
line upto $J=5$. $c$ is on the horizontal-axis.}\label{j34.fig} 
%%%%%%%%%%%%%%%%%%%%%%%%%%%%%%%%%%%%%

%\begin{figure}[ht]
%\centering
%\resizebox{3.5in}{!}{\includegraphics{100-3.eps}}
%\caption{ The red line is the sum $\U_{3,4}$ upto $J=5$ and the dashed black
%line upto $J=100$. $c$ is on the horizontal-axis.}\label{j34.fig} 
%\end{figure}

\bibliographystyle{unsrt}
\bibliography{refs}
\end{document}